\documentclass[10pt,conference]{IEEEtran}

\IEEEoverridecommandlockouts
\usepackage{bbm}
\usepackage{amsfonts}
\usepackage[dvips]{graphicx}
\usepackage{times}
\usepackage{cite}
\usepackage{amsmath}
\usepackage{array}
\usepackage{amssymb}

\usepackage{stfloats}
\usepackage{graphicx}
\usepackage{footnote}
\usepackage{booktabs}
\usepackage{array}
\usepackage{algorithmic}
\usepackage{algorithm}
\usepackage{subeqnarray}
\usepackage{cases}
\usepackage{threeparttable}
\usepackage{color}
\usepackage{epstopdf}
\usepackage{cleveref}
\usepackage{bm}
\usepackage{subcaption}
\usepackage{dsfont}

\newtheorem{proof}{Proof}

\begin{document}

\title{Exploiting Tradeoff Between Transmission Diversity and Content Diversity in Multi-Cell Edge Caching}
\author{\IEEEauthorblockN{Kangqi Liu and Meixia Tao}
\IEEEauthorblockA{Department of Electronic Engineering, Shanghai Jiao Tong University, Shanghai, China\\
\thanks{This work is supported by the National Natural Science Foundation of China under grants 61571299 and 61329101.}
Emails: k.liu.cn@ieee.org, mxtao@sjtu.edu.cn}
}
\maketitle

\begin{abstract}
Caching in multi-cell networks faces a well-known dilemma, i.e., to cache same contents among multiple edge nodes (ENs) to enable transmission cooperation/diversity for higher transmission efficiency, or to cache different contents to enable content diversity for higher cache hit rate. In this work, we introduce a partition-based caching to exploit the tradeoff between transmission diversity and content diversity in a multi-cell edge caching networks with single user only. The performance is characterized by the system average outage probability, which can be viewed as the sum of the cache hit outage probability and cache miss probability. We show that (i) In the low signal-to-noise ratio(SNR) region, the ENs are encouraged to cache more fractions of the most popular files so as to better exploit the transmission diversity for the most popular content; (ii) In the high SNR region, the ENs are encouraged to cache more files with less fractions of each so as to better exploit the content diversity.
\end{abstract}

\section{Introduction}
Caching can alleviate peak-hour network congestion, provide traffic offloading, and improve users' quality of experience by prefetching popular contents during off-peak times at the edge of wireless networks, such as base stations and user devices \cite{Maddah-Ali,Golrezaei,Bastug,Liu10}. Caching, at the same time, introduces new challenges for system design since the cache placement needs to be jointly considered with the content delivery by taking into account the wireless aspects of the network.

The caching design in multi-cell networks faces a well-known dilemma, i.e., to cache same contents or to cache different contents in the multiple edge nodes (ENs). To cache the same contents, one can achieve the physical-layer transmission cooperation (or diversity) gain and hence improve spectral efficiency and transmission reliability. To cache different contents, one can achieve the content diversity gain and hence increase the cache hits, thereby reducing backhaul traffic. Recently, much attention has been drawn to balancing these two gains \cite{Tao2,Ao,Chen2,Song,Chae,Xu3,Maddah-Ali3,Xu2,Xu,Cao,Cao2}. In \cite{Tao2}, three heuristic caching schemes together with content-centric sparse multicast beamforming are proposed to  balance the system total energy consumption and backhaul consumption in a cache-enabled cloud radio access network. It is shown with simulation that caching the most popular contents in each EN generally outperforms probabilistic caching for large user density. In \cite{Ao}, a small cell cooperation with threshold-based caching method is proposed to combine the advantages of distributed caching and physical layer cooperative transmission. In \cite{Chen2}, the tradeoff between transmission cooperation gain and content diversity gain is investigated based on a caching scheme where each file is either cached at all the ENs entirely or equally split into subfiles and cached in each EN without overlapping. In \cite{Xu3}, the authors studied the coded caching with maximum distance separable codes or random linear network codes in small-cell networks. It is shown that coded caching outperforms the most popular caching (MPC) strategy in terms of the average fractional offloaded traffic and average ergodic rate when the content popularity skewness parameter is small. Yet, this coded caching cannot exploit the transmission cooperation (or diversity) gain since each base station caches independent coded packets and each user adopts successive interference cancellation-based receiver. In \cite{Maddah-Ali3}, the authors studied the transmission cooperation gain in a $3\times 3$ interference channel with cache equipped at transmitters. In \cite{Xu2,Xu,Cao,Cao2}, the tradeoff between storage and latency in interference networks with caches equipped at both the transmitter side and the receiver side are investigated. However, it is assumed in \cite{Xu2,Xu,Cao,Cao2} that the accumulated cache size at all the nodes (including transmitters and receivers) are large enough  to collectively store the entire content library without the need for backhaul.

\begin{figure}[t]
\begin{centering}
\includegraphics[scale=0.2]{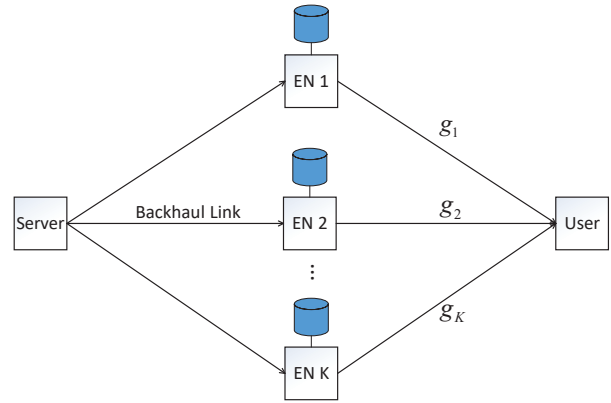}
\vspace{-0.1cm}
 \caption{Multi-cell network with edge caching.}\label{Fig_model}
\end{centering}
\vspace{-0.3cm}
\end{figure}

This work aims to exploit the tradeoff between transmission diversity and content diversity via partition-based caching in a multi-cell network with edge caching as shown in Fig. \ref{Fig_model}, where for simplicity, only one user is considered. Our proposed caching strategy allows that each file can be split into multiple subfiles and each subfile can be cached in multiple ENs. This enables partial ENs to transmit the subfiles cooperatively, which is called as \textit{partial transmission cooperation gain}. It is seen that there is a tradeoff between the cache hit probability and transmission cooperation gain. We study this tradeoff by minimizing the average outage probability of the system. Our study shows that: (i) In the low SNR region, the ENs are encouraged to cache more fractions of the most popular files so as to better exploit the transmission diversity for the most popular content; (ii) In the high SNR region, the ENs are encouraged to cache more files with less fractions of each so as to better exploit the content diversity. Our numerical results show that our considered partial transmission cooperation scheme has a better performance than the full transmission cooperation scheme in \cite{Chen2}.

\section{System Model}
\subsection{Channel Model}
Consider a multi-cell caching network as shown in Fig. \ref{Fig_model}, where there are $K$ cache-enabled ENs and one user. Each EN is connected to the server via a dedicated backhaul link. Each node is assumed to have single antenna. The file library consists of a set of $N$ files, denoted by ${\cal W}=\{W_1,W_2,\cdots,W_N\}$. Each file has equal length of $F$ bits, where $F$ is assumed to be large enough. Each EN is assumed to have a local cache of size $MF$ bits ($M$ files) with $M < N$. The user requests a file from the file library according to a known probability distribution ${\bf p}=(p_1,p_2,\cdots,p_N)$, i.e., $W_i$ is requested with probability $p_i$. In this paper, the file popularity is modeled as the Zipf distribution with skewness parameter $\rho$ following the convention in \cite{Bastug2,Shanmugam,Golrezaei}, i.e.,
\begin{align}
p_i=\frac{i^{-\rho}}{\sum_{n=1}^{N} n^{-\rho}},~1\leq i \leq N.
\end{align}

The communication takes place in two phases, a cache placement phase and a content delivery phase. During the cache placement phase, each EN fills up its cache memory via a backhaul link with some contents from the file library without having the knowledge of user request and channel state information. In the content delivery phase, the user requests a file $W_d$ from ${\cal W}$, where $d \in \{1,2,\cdots, N\}$ with probability $\mathbb{P} (d=i)=p_i$. The channel gain, denoted by $g_k$, from the $k$-th EN to the user follows the distribution ${\cal CN}(0,1)$, and $\{g_k\}_{k=1}^K$ are independent of each other. The average transmission power of each EN is $P$. Each EN-user link experiences an additive white Gaussian noise with distribution ${\cal CN}(0,1)$.

\subsection{Caching Model}
We proposed a partition-based caching scheme. The cache memory at each EN is split into $N_0 \leq N$ parts, with each part of size $\mu_i F$ bits and used to cache $\mu_i F$ bits of file $W_i$, for $1\leq i \leq N_0$. Here, $N_0$ and $\mu_i$ are all design variables satisfying the following constraint:
\begin{align}
&\sum_{i=1}^{N_0} \mu_i \leq M,\\
&0<\mu_i \leq 1,\\
&1\leq N_0 \leq N,~N_0 \in \mathbb{N}.
\end{align}
In this paper, we only consider that $K\mu_i$ are all integers. Define
\begin{align}
t_i \triangleq K\mu_i,~\forall i\in \{1,2, \cdots, N_0\}.
\end{align}
Here, $t_i$ can be viewed as the total cache size (normalized by file size $F$) in the network edge allocated to file $W_i$. For each file $W_i$, $i=1,2,\cdots,N_0$, it is partitioned into $\binom{K}{t_i}$ disjoint subfiles of equal size, denoted by
\begin{align}
W_i=\{W_{i,{\cal T}}:{\cal T} \subseteq \{1,2,\cdots,K\},|{\cal T}|=t_i\}.
\end{align}
Each subfile $W_{i,{\cal T}}$ is cached in all the $t_i$ ENs in set ${\cal T}$. Thus, every bit of each file $W_i$ is cached simultaneously at $t_i$ distinct ENs. Based on this partitioning, each EN $k$, for $k=1,2,\cdots,K$, stores a set of subfiles in the library denoted by
\begin{align}
{\cal Z}_k=\{W_{i,{\cal T}}\mid k \in {\cal T}\}_{i=1}^{N_0}
\end{align}

\textit{Remark 1:} The caching scheme in \cite{Chen2} only considered the special case with $t_i=1$ $(\mu_i=\frac{1}{K})$ or $t_i=K$ $(\mu_i=1)$. In this paper, we consider the more general case where $t_i$ (or equivalently $\mu_i$) can be optimized within $[1,K]$.

\subsection{Delivery Model and Performance Metric}
In the content delivery phase, assume that the file $W_d$ is requested. The delivery scheme is elaborated as follows.

\subsubsection{$1\leq d \leq N_0$ (cache hit)}
For this case, the file $W_d$ is cached by the ENs. Each subfile $W_{d,{\cal T}}$ is transmitted by its associated set of caching ENs ${\cal T}$ cooperatively and different subfiles are transmitted sequentially over the time independent to each other. The channel thus becomes a $t_d \times 1$ multiple-intput single-output (MISO) channel. The capacity of this MISO channel is given by
\begin{align}\label{rate_set}
R_{d,{\cal T}}=\log_2\left(1+\sum_{k \in {\cal T}}|g_k|^2 P\right),
\end{align}

The transmission of file $W_d$ is said to be in outage if the transmission rate of any subfile $W_{d,{\cal T}}$ falls below a target rate, denoted by $R$. By this definition, the outage probability of transmitting file $W_d$, $1\leq d \leq N_0$, is given by
\begin{align}\label{P_out_set_min}
P_{\textrm{out}}(d,P,R)=\mathbb{P}\left(\min_{{\cal T} \subseteq \{1,2,\cdots,K\},|{\cal T}|=t_d} \{R_{d,{\cal T}}\}<R\right).
\end{align}
The transmission diversity of file $W_d$, $1\leq d \leq N_0$, is given by
\begin{align}
\gamma(d)\triangleq \lim_{P \to \infty}-\frac{\log (P_{out}(d,P,R))}{\log P}.
\end{align}

\subsubsection{$N_0 < d \leq N$ (cache miss)}
For this case, the ENs will fetch the content from the backhaul link and then transmit to the user. To simplify our analysis, we do not consider any specific delivery scheme for the cache miss case. We only treat this as a transmission outage (from the local cache of ENs) for file $W_d$. That is, the transmission of $W_d$ is in outage if it is not cached. The transmission diversity of file $W_d$, $N_0 < d \leq N$, is 0.

\subsubsection{System performance metrics}
After defining the outage probability for each file request, the average outage probability of the system is given by
\begin{align}\label{PM}
P_{\textrm{out}}^{s}(P,R)\triangleq& \sum_{d=1}^{N_0} p_d P_{\textrm{out}}(d,P,R)+\sum_{d=N_0+1}^{N} p_d.
\end{align}
The transmission diversity of the system is defined as
\begin{align}
\gamma^s\triangleq & \lim_{P \to \infty} -\frac{\log (P_{\textrm{out}}^{s}(P,R))}{\log P}=\min_{1 \leq i \leq N} \{\gamma(d_i)\}.
\end{align}
The transmission diversity for cache hit is defined as
\begin{align}
\gamma^{hit}\triangleq & \lim_{P \to \infty} -\frac{\sum_{d=1}^{N_0} p_d P_{\textrm{out}}(d,P,R)}{\log P}=\min_{1 \leq i \leq N_0} \{\gamma(d_i)\}.
\end{align}

The content diversity is defined as $N_0$.

It is seen from \eqref{PM} that the performance metric can be viewed as the sum of the cache hit transmission outage probability $\sum_{d=1}^{N_0} p_d P_{\textrm{out}}(d,P,R)$ and the cache miss probability $\sum_{d=N_0+1}^{N} p_d$. With the increase of the content diversity $N_0$, the transmission diversity $\gamma^{hit}$ decreases because $t_d$ decreases. Thus, the second term in \eqref{PM} becomes smaller while the first term in \eqref{PM} becomes larger. The average outage probability of the system is able to characterize the tradeoff between transmission diversity and content diversity.

\section{Outage Probability Analysis ($1 \leq d \leq N_0$)}
In this section, we present the outage probability of the file request $1 \leq d \leq N_0$.

\textit{Theorem 1:} When file $W_d$ ($1 \leq d \leq N_0$) is requested, the outage probability is given by \eqref{TH1} on top of this page with
\begin{figure*}
\begin{align}\label{TH1}
&P_{\textrm{out}}(d,P,R)=\left\{
                      \begin{array}{ll}
                        1-\sum\limits_{i=1}^{t_d} \frac{(c_i)^{t_d-1}}{\prod_{j=1,j\neq i}^{t_d} (c_i-c_j)}e^{-\frac{2^{R}-1}{Pc_i}}, & \hbox{$1\leq t_d <K$;} \\
                        1-\sum\limits_{i=1}^{K-1}\frac{\left(\frac{2^{R}-1}{P}\right)^i}{i!}e^{-\frac{2^{R}-1}{P}}, & \hbox{$t_d=K$.}
                      \end{array}
                    \right.
\end{align}
\hrule
\vspace{-0.7cm}
\end{figure*}
$c_i=\frac{t_d-i+1}{K-i+1}$, and the transmission diversity is $\gamma(d) = t_d$.

The diversity result in this theorem is intuitive since every bit of file $W_d$ is cached in $t_d$ distinct ENs. However, the proof of the outage probability is rather challenging due to the correlation of each rate expression $R_{d,{\cal T}}$ in \eqref{P_out_set_min} when $|{\cal T}|>1$.

\subsection{Proof of Outage probability}
By definition in \eqref{P_out_set_min}, we have
\begin{subequations}\label{Outage_Probability_1}
\begin{align}\nonumber
&P_{\textrm{out}}(d,P,R)\\\label{Outage_Probability_1_a}
=&\mathbb{P}\left(\min_{\forall {\cal T},|{\cal T}|=t_d} \log_2\left(1+\sum_{k \in {\cal T}}|g_k|^2 P\right)<R\right)\\\label{Outage_Probability_1_b}
=&\mathbb{P}\left(\log_2\left(1+\min_{\forall {\cal T},|{\cal T}|=t_d} \sum_{k \in {\cal T}}|g_k|^2 P\right)<R\right)\\\label{Outage_Probability_1_c}
=&\mathbb{P}\left(\min_{\forall {\cal T},|{\cal T}|=t_d} \sum_{k \in {\cal T}}|g_k|^2<\frac{2^R-1}{P}\right).
\end{align}
\end{subequations}
For simplicity of presentation, we define $X_k\triangleq |g_k|^2$. Since $\{g_k\}_{k=1}^K$ are independent and identically distributed with ${\cal CN}(0,1)$, $\{X_k\}_{k=1}^K$ are $K$ statistically independent variables with standard exponential distribution. Let $X_{(1)} \leq X_{(2)} \leq \cdots \leq X_{(K)}$ denote the ordered sample $\{X_k\}_{k=1}^K$. Then, \eqref{Outage_Probability_1_c} can be rewritten as
\begin{align}\label{Outage_Probability_1_d}
P_{\textrm{out}}(d,P,R)&=\mathbb{P}\left(\sum_{k=1}^{t_d} X_{(k)}<\frac{2^R-1}{P}\right).
\end{align}
Note that $\{X_{(k)}\}_{k=1}^K$ are no longer independent of each other , making the distribution of $\sum_{k=1}^{t_d} X_{(k)}$ difficult to obtain. Thus, we introduce the following lemma first.

\textit{Lemma 1:} Let $X_{(1)} \leq X_{(2)} \leq \cdots \leq X_{(K)}$ denote the ordered variables in a sample of $K$ from the standard exponential distribution, then $X_{(k)}$ can be expressed as
\begin{align}\label{transform}
X_{(k)}=\sum_{i=1}^k \frac{Z_i}{K-i+1},
\end{align}
where $\{Z_i\}_{i=1}^K$ are $K$ statistically independent variables with standard exponential distribution.
\begin{proof}
See Appendix A.
\end{proof}

With \textit{Lemma 1}, we can express $\sum_{k=1}^{t_d} X_{(k)}$ as
\begin{subequations}\label{transform2}
\begin{align}\label{transform2_a}
\sum_{k=1}^{t_d} X_{(k)}=&\sum_{k=1}^{t_d} \sum_{i=1}^k \frac{Z_i}{K-i+1}\\\label{transform2_b}
=&\sum_{k=1}^{t_d} \underbrace{\frac{t_d-k+1}{K-k+1}}_{\triangleq c_k} Z_k\\\label{transform2_c}
=&\sum_{k=1}^{t_d} c_k Z_k
\end{align}
\end{subequations}
Thus far, it is seen that the calculation of \eqref{Outage_Probability_1_d} is equivalent to showing the distribution of a linear combination of standard exponential variables without order. To proceed the analysis, we present a useful lemma in \cite{Ali} as below:

\textit{Lemma 2 (\cite[Theorem 3]{Ali}):} Let $\{Z_i\}_{i=1}^n$ be independent and identically distributed standard exponential random variables. Then, the CDF of $\sum_{i=1}^{n} c_i Z_i$ is given by
\begin{align}\label{CDF_1}
\mathbb{P} \left(\sum_{i=1}^{n} c_i Z_i <T\right)=1-\left[(z_{+})^K e^{-\frac{T}{z}} \mid z=c_0,c_1,\cdots,c_{K}\right],
\end{align}
where $x_{+}\triangleq\max\{x,0\}$, $c_0\triangleq0$ and $[f(x) \mid x=x_0,x_1,\cdots,x_r]$ represents the $r$-th divided difference of a function $f(x)$ with arguments $x=x_0,x_1,\cdots,x_r$\footnote{For the calculation of divided difference, we refer interested readers to Appendix B for details.}.

When $1\leq t_d <K$, it can be seen from \eqref{transform2_b} that $c_0,c_1,\cdots,c_{t_d}$ are distinct. Thus, the outage probability can be calculated from \eqref{transform2}\eqref{CDF_1}\eqref{divided_difference}.

When $t_d=K$, $c_0,c_1,\cdots,c_{t_d}$ are all equal to 1 from \eqref{transform2_b}. The calculation of the divided difference is complicated. Thus, we obtain the distribution of this special case directly by finding the distribution of the sum of $K$ standard exponential variables. From the property of the Gamma distribution, $\sum_{k=1}^{K} X_{k}$ follows Gamma distribution with shape $K$ and scale $1$. The first part of the theorem is proved.

\subsection{Proof of transmission diversity}
We first consider the case $1\leq t_d <K$. For the simplicity of presentation, we define
\begin{align}\label{T_1}
T_1\triangleq \frac{2^R-1}{P}.
\end{align}
By using the series expansion
\begin{align}\label{series_expansion}
e^x=\sum_{m=0}^{\infty} \frac{x^m}{m!},
\end{align}
the outage probability in \eqref{TH1} can be rewritten as
\begin{subequations}\label{P_out_1}
\begin{align}\nonumber
&P_{\textrm{out}}(d,P,R)\\
=&1-\sum\limits_{i=1}^{t_d} \frac{(c_i)^{t_d-1}}{\prod_{j=1,j\neq i}^{t_d} (c_i-c_j)}e^{-\frac{2^{R}-1}{Pc_i}}\\
=&1-\sum_{i=1}^{t_d} \left[\frac{(c_i)^{t_d-1}}{\prod_{j=1,j\neq i}^{t_d} (c_i-c_j)} \sum_{m=0}^{\infty} \frac{(-\frac{T_1}{c_i})^m}{m!}\right]\\
=&1-\sum_{i=1}^{t_d} \sum_{m=0}^{\infty} \left[\frac{(c_i)^{t_d-1-m}}{\prod_{j=1,j\neq i}^{t_d} (c_i-c_j)} \frac{T_1^m (-1)^m}{m!}\right]\\
=&1-\sum_{m=0}^{\infty} \sum_{i=1}^{t_d} \left[\frac{(c_i)^{t_d-1-m}}{\prod_{j=1,j\neq i}^{t_d} (c_i-c_j)} \frac{(-1)^m}{m!}T_1^m\right]\\
=&1-\sum_{m=0}^{\infty} \left[T_1^m \underbrace{\sum_{i=1}^{t_d}\frac{(c_i)^{t_d-1-m}}{\prod_{j=1,j\neq i}^{t_d} (c_i-c_j)} \frac{(-1)^m}{m!}}_{\triangleq f(t_d,m)}\right].
\end{align}
\end{subequations}

To proceed the analysis, we introduce the following lemma.

\textit{Lemma 3:} For given $K$ and $t_d$, we have
\begin{align}
f(t_d,m)=\left\{
      \begin{array}{ll}
        1, & \hbox{$m=0$;} \\
        0, & \hbox{$1 \leq m \leq t_d-1$.}
      \end{array}
    \right.
\end{align}
and
\begin{align}
f(t_d,m=t_d)\neq 0.
\end{align}

\begin{proof}
See Appendix C.
\end{proof}

With \textit{Lemma 3}, \eqref{P_out_1} can be rewritten as
\begin{align}
P_{\textrm{out}}(d,P,R)=&-\sum_{m=t_d}^{\infty} f(t_d,m)T_1^m.
\end{align}
The transmission diversity can be calculated as
\begin{subequations}
\begin{align}
\gamma(d)=& \lim_{P \to \infty} -\frac{\log (P_{\textrm{out}}(d,P,R))}{\log P}\\
=&\lim_{P \to \infty} -\frac{\log \left(\sum_{m=t_d}^{\infty} -f(t_d,m)T_1^m\right)}{\log P}\\
=&\lim_{P \to \infty} -\frac{\log \left(\sum_{m=t_d}^{\infty} -f(t_d,m)\left(2^R-1\right)^mP^{-m}\right)}{\log P}\\
=&t_d
\end{align}
\end{subequations}

Next, we consider the case $t_d=K$. With \eqref{T_1} and \eqref{series_expansion}, the outage probability in \eqref{TH1} can be rewritten as
\begin{subequations}\label{P_out_2}
\begin{align}\label{P_out_2_a}
P_{out}(d,P,R)=&1-\sum\limits_{i=1}^{K-1}\frac{T_1^i}{i!}e^{-\frac{2^{R}-1}{P}}\\\label{P_out_2_b}
=&e^{T_1}e^{-T_1}-\sum\limits_{i=1}^{K-1}\frac{T_1^i}{i!}e^{-T_1}\\\label{P_out_2_c}
=&\sum\limits_{i=1}^{\infty}\frac{T_1^i}{i!}e^{-T_1}-\sum\limits_{i=1}^{K-1}\frac{T_1^i}{i!}e^{-T_1}\\\label{P_out_2_d}
=&\sum\limits_{i=K}^{\infty}\frac{T_1^i}{i!}e^{-T_1}\\\label{P_out_2_e}
=&\sum\limits_{i=K}^{\infty}\sum\limits_{m=0}^{\infty}\frac{T_1^{i+m} (-1)^m}{i!m!}
\end{align}
\end{subequations}
The transmission diversity can be calculated as
\begin{subequations}
\begin{align}
\gamma(d)=& \lim_{P \to \infty} -\frac{\log (P_{out}(d,P,R))}{\log P}\\
=&\lim_{P \to \infty} -\frac{\log \left(\sum\limits_{i=K}^{\infty}\sum\limits_{m=0}^{\infty}\frac{T_1^{i+m} (-1)^m}{i!m!}\right)}{\log P}\\
=&\lim_{P \to \infty} -\frac{\log \left(\sum\limits_{i=K}^{\infty}\sum\limits_{m=0}^{\infty}\frac{\left(2^R-1\right)^{i+m} P^{-i-m} (-1)^m}{i!m!}\right)}{\log P}\\
=&K
\end{align}
\end{subequations}
Thus far, \textit{Theorem 1} is proved.

\section{Minimum Outage Probability of the System}
In the previous sections, we have obtained the outage probability for each file request for given design parameters $N_0$ and $t_i$ $(\mu_i)$. The average outage probability of the system can be minimized by optimizing $N_0$ and $\{t_i\}_{i=1}^{N_0}$. This is formulated as follows.
\begin{subequations}\label{OP}
\begin{align}\nonumber
&\min_{N_0,\{t_i\}_{i=1}^{N_0}}~\sum_{d=1}^{N_0} p_d P_{\textrm{out}}(d,P,R)+\sum_{d=N_0+1}^{N} p_d\\
s.t.~~& 1\leq N_0 \leq N,~N_0 \in \mathbb{N},\\
&1\leq t_i \leq K,~t_i\in \mathbb{N},~1 \leq i \leq N_0,\\
&\sum_{i=1}^{N_0} \frac{t_i}{K} \leq M.
\end{align}
\end{subequations}
Note that the above problem is an integer programming problem. A brute-force approach can be used to find the global optimal solution. In particular, there are at most $K^{N}$ possible choices of $\{N_0,\{t_i\}_{i=1}^{N_0}\}$. The search space can be reduced by exploiting the optimality condition that $t_1 \geq t_2 \geq \cdots \geq t_{N_0}$. Nevertheless, it is worth mentioning that the cache placement occurs in the off-peak time and hence the optimization problem can be solved off-line.

Next, we present the tradeoff between the transmission diversity for cache hit $\gamma^{hit}$ and the content diversity $N_0$ for each file under uniform file demand, i.e., $p_d=\frac{1}{K}$ for all $d$. For this case, it is seen from \eqref{OP} that the optimal $t_i$, $1\leq i \leq N_0$, are the same due to the symmetry of each file. Thus, $\gamma^{hit}$ can be expressed as
\begin{align}
\gamma^{hit}=\min\{t_1,t_2,\cdots,t_{N_0}\}=\min\left\{\frac{MK}{N_0},K\right\}.
\end{align}
When $N_0 \leq M$, this means the optimal file splitting is to cache all the first $N_0$ files at all the ENs. When $N_0 > M$, it is seen that $\gamma^{hit}=\frac{MK}{N_0}$, which indicates $\gamma^{hit}$ is inversely proportional to $N_0$.

\section{Numerical Results}
\begin{figure}[t]
\begin{centering}
\includegraphics[scale=0.35]{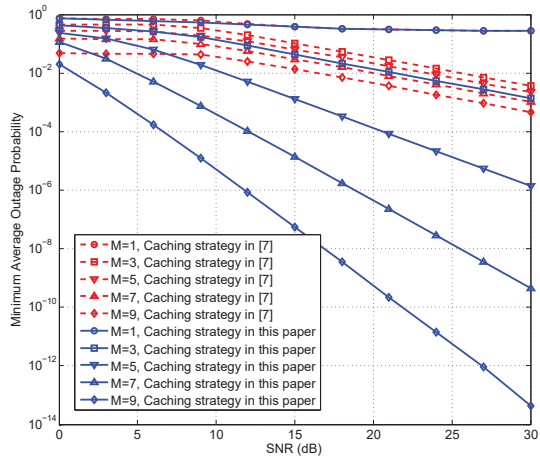}
\vspace{-0.1cm}
 \caption{Outage probability performance of the system.}\label{Fig_New_Outage_K_5_M_1_3_5_7_9_N_10_ours_vs_Tony}
\end{centering}
\vspace{-0.3cm}
\end{figure}

\begin{table}[]
\tiny
\centering
\caption{Optimal selection of $\{N_0,\{t_i\}_{i=1}^{N_0}\}$ of our proposed scheme when $K=5$ and $N=10$}
\label{table_1}
\begin{tabular}{|c|c|c|c|c|c|c|c|c|c|c|c|}
\hline
$M=1$       & $N_0$ & $t_1$ & $t_2$ & $t_3$ & $t_4$ & $t_5$ & $t_6$ & $t_7$ & $t_8$ & $t_9$ & $t_{10}$ \\ \hline
0dB         & 1    & 5    & 0    & 0    & 0    & 0    & 0    & 0    & 0    & 0    & 0         \\ \hline
3dB-6dB     & 2    & 3    & 2    & 0    & 0    & 0    & 0    & 0    & 0    & 0    & 0         \\ \hline
9dB         & 3    & 2    & 2    & 1    & 0    & 0    & 0    & 0    & 0    & 0    & 0         \\ \hline
12dB        & 4    & 2    & 1    & 1    & 1    & 0    & 0    & 0    & 0    & 0    & 0         \\ \hline
15dB-30dB   & 5    & 1    & 1    & 1    & 1    & 1    & 0    & 0    & 0    & 0    & 0         \\ \hline
$M=3$       & $N_0$ & $t_1$ & $t_2$ & $t_3$ & $t_4$ & $t_5$ & $t_6$ & $t_7$ & $t_8$ & $t_9$ & $t_{10}$ \\ \hline
0dB       & 4    & 4    & 4    & 4    & 3    & 0    & 0    & 0    & 0    & 0    & 0         \\ \hline
3dB       & 5    & 3    & 3    & 3    & 3    & 3    & 0    & 0    & 0    & 0    & 0         \\ \hline
6dB       & 7    & 3    & 2    & 2    & 2    & 2    & 2    & 2    & 0    & 0    & 0         \\ \hline
9dB       & 9    & 2    & 2    & 2    & 2    & 2    & 2    & 1    & 1    & 1    & 0         \\ \hline
12dB-30dB & 10    & 2    & 2    & 2    & 2    & 2    & 1    & 1    & 1    & 1    & 1         \\ \hline
$M=5$       & $N_0$ & $t_1$ & $t_2$ & $t_3$ & $t_4$ & $t_5$ & $t_6$ & $t_7$ & $t_8$ & $t_9$ & $t_{10}$ \\ \hline
0dB       & 6    & 5    & 4    & 4    & 4    & 4    & 4    & 0    & 0    & 0    & 0         \\ \hline
3dB       & 8    & 4    & 3    & 3    & 3    & 3    & 3    & 3    & 3    & 0    & 0         \\ \hline
6dB-30dB  & 10   & 3    & 3    & 3    & 3    & 3    & 2    & 2    & 2    & 2    & 2         \\ \hline
$M=7$       & $N_0$ & $t_1$ & $t_2$ & $t_3$ & $t_4$ & $t_5$ & $t_6$ & $t_7$ & $t_8$ & $t_9$ & $t_{10}$ \\ \hline
0dB       & 8    & 5    & 5    & 5    & 4    & 4    & 4    & 4    & 4    & 0    & 0         \\ \hline
3dB-30dB  & 10   & 4    & 4    & 4    & 4    & 4    & 3    & 3    & 3    & 3    & 3         \\ \hline
$M=9$       & $N_0$ & $t_1$ & $t_2$ & $t_3$ & $t_4$ & $t_5$ & $t_6$ & $t_7$ & $t_8$ & $t_9$ & $t_{10}$ \\ \hline
0dB-30dB  & 10   & 5    & 5    & 5    & 5    & 5    & 4    & 4    & 4    & 4    & 4         \\ \hline
\end{tabular}
\end{table}

In this section, we provide numerical results to show the performance of the system. The target data rate $R$ for each message is set to be 1 bit/s/Hz. The file popularity is modeled as the Zipf distribution with parameter $0.8$. We consider the scenario where there are $K=5$ ENs and $N=10$ files in the library. The outage probability performance compared with \cite{Chen2} is illustrated in Fig. \ref{Fig_New_Outage_K_5_M_1_3_5_7_9_N_10_ours_vs_Tony}. When $M=1$ and $M=3$, i.e., the cache size of each EN is small, it is seen from Fig. \ref{Fig_New_Outage_K_5_M_1_3_5_7_9_N_10_ours_vs_Tony} that the transmission diversity of the system are the same with the two strategies, i.e., $\gamma^s=0$ when $M=1$ and $\gamma^s=1$ when $M=3$. However, when $M$ becomes larger, the transmission diversity of the system of our proposed caching strategy increases ($\gamma^s=2$ when $M=5$, $\gamma^s=3$ when $M=7$, and $\gamma^s=4$ when $M=9$) while that of the strategy proposed in \cite{Chen2} remains to be 1. This is because the authors in \cite{Chen2} only considered the full transmission cooperation of all the ENs, which occupies too much storage at the EN for each file. Our proposed strategy consider the partial transmission cooperation. The storage occupied at the EN for each file is less than that of the full transmission cooperation scheme, which can better exploit both transmission diversity and content diversity. Furthermore, it is seen that the outage probability of the scheme proposed in \cite{Chen2} remains the same in the low SNR region while the outage probability of our proposed scheme decreases when SNR increases in the low SNR region.

The optimal selection of $\{N_0,\{t_i\}_{i=1}^{N_0}\}$ of our proposed scheme is given in TABLE \ref{table_1}. In the low SNR region, it is seen from TABLE \ref{table_1} that $N_0$ is small such that we can select a large value for each nonzero $t_i$ to better exploit the transmission diversity for the most popular content. In the high SNR region, it is seen from TABLE \ref{table_1} that $N_0$ is large such that more content can be cached in the EN to increase the cache hit rate such that the content diversity of the system increases.
\section{Conclusion}
In this paper, we studied the tradeoff between transmission diversity and content diversity in a multi-cell network with edge caching. We proposed a partition-based caching scheme and a partial transmission cooperation delivery scheme which can exploit both transmission diversity and content diversity. We obtained two main results for this tradeoff. In the low SNR region, the ENs are encouraged to cache more fractions of the most popular files so as to better exploit the transmission diversity for the most popular content. In the high SNR region, the ENs are encouraged to cache more files with less fractions of each so as to better exploit the content diversity. In the future work, we are interested in studying the tradeoff between transmission diversity and content diversity in more complicated cache-aided wireless networks.

\section*{Appendix A: Proof of \textit{Lemma 1}}
The joint PDF of $\{X_{(i)}\}_{i=1}^K$ is given by \cite{David}
\begin{align}
K!\exp\left(-\sum_{i=1}^K x_{(i)}\right),~0\leq x_{(1)} \leq x_{(2)} \leq \cdots \leq x_{(K)} \leq +\infty,
\end{align}
which can be rewritten as
\begin{align}\label{joint_pdf_transform}
K!\exp\left(-\sum_{i=1}^K (K-i+1)(x_{(i)}-x_{(i-1)})\right),
\end{align}
where $x_{(0)}=0$. Define
\begin{align}\label{z_i}
z_i \triangleq (K-i+1)(x_{(i)}-x_{(i-1)}).
\end{align}
It is seen that $\{Z_i\}_{i=1}^K$ are $K$ statistically independent variables with standard exponential distribution \cite{David}. From \eqref{z_i}, $X_{(k)}$ can be expressed as
\begin{align}
X_{(k)}=\sum_{i=1}^k (X_{(i)}-X_{(i-1)})=\sum_{i=1}^k \frac{Z_i}{K-i+1}.
\end{align}
The lemma is thus proved.

\section*{Appendix B: Calculation of divided difference}
When $x_0,x_1,\cdots,x_r$ are distinct, the divided difference can be calculated as \cite{Hildebrand,Ali1}
\begin{align}\label{divided_difference}
\left[f(x) \mid x=x_0,x_1,\cdots,x_r\right]=&\sum_{i=0}^r \frac{f(x_i)}{\prod_{j=0,j\neq i}^r (x_i-x_j)}=\frac{|{\bf A}|}{|{\bf B}|}.
\end{align}
where
\begin{align}
{\bf A}=\left[
          \begin{array}{cccccc}
            1 & c_0 & c_0^2 & \cdots & c_0^{t_d-1} & f(c_0)\\
            1 & c_1 & c_1^2 & \cdots & c_1^{t_d-1} & f(c_1)\\
            1 & c_2 & c_2^2 & \cdots & c_2^{t_d-1} & f(c_2)\\
            \vdots & \vdots & \vdots & \ddots & \vdots & \vdots\\
            1 & c_{t_d} & c_{t_d}^2 & \cdots & c_{t_d}^{t_d-1} & f(c_{t_d})\\
          \end{array}
        \right]
\end{align}
\begin{align}
{\bf B}=\left[
          \begin{array}{ccccc}
            1 & c_0 & c_0^2 & \cdots & c_0^{t_d} \\
            1 & c_1 & c_1^2 & \cdots & c_1^{t_d} \\
            1 & c_2 & c_2^2 & \cdots & c_2^{t_d} \\
            \vdots & \vdots & \vdots & \ddots & \vdots \\
            1 & c_{t_d} & c_{t_d}^2 & \cdots & c_{t_d}^{t_d} \\
          \end{array}
        \right]
\end{align}
For coincident value of $x_0,x_1,\cdots,x_r$, the calculation of the divided difference is much more complicated and thus not presented here.

\section*{Appendix C: Proof of \textit{Lemma 3}}
From \eqref{divided_difference}, we can rewrite $f(t_d,m)$ as
\begin{subequations}
\begin{align}
f(t_d,m)=&\frac{(-1)^m}{m!}\sum_{i=1}^{t_d} \frac{(c_i)^{t_d-1-m}}{\prod_{j=1,j\neq i}^{t_d} (c_i-c_j)}\\
=&\frac{(-1)^m}{m!}\left[(c)^{t_d-1-m} \mid c=c_1,c_2,\cdots, c_{t_d}\right]\\
=&\frac{(-1)^m}{m!}\frac{|{\bf A}|}{|{\bf B}|},
\end{align}
\end{subequations}
where $f(c_i)=c_i^{t_d-1-m}$.

When $m=0$, it is seen that ${\bf A}={\bf B}$ and thus we have $f(t_d,0)=1$. When $1 \leq m \leq t_d$, it is seen that the last column of $\bf A$ is the same as the last $m+1$ column of $\bf A$ when $1 \leq m \leq t_d-1$, which indicates $|{\bf A}|=0$ when $1 \leq m \leq t_d-1$. Thus, $f(t_d,m)=0$ when $1 \leq m \leq t_d-1$. When $m=t_d$, it is seen that $\bf A$ is a full-rank matrix and ${\bf A} \neq {\bf B}$ and thus $f(t_d,m)\neq 0$ with probability 1.

Thus far, the lemma is proved.

\bibliographystyle{IEEEtran}
\bibliography{IEEEabrv,reference}

\begin{thebibliography}{10}
\providecommand{\url}[1]{#1}
\csname url@samestyle\endcsname
\providecommand{\newblock}{\relax}
\providecommand{\bibinfo}[2]{#2}
\providecommand{\BIBentrySTDinterwordspacing}{\spaceskip=0pt\relax}
\providecommand{\BIBentryALTinterwordstretchfactor}{4}
\providecommand{\BIBentryALTinterwordspacing}{\spaceskip=\fontdimen2\font plus
\BIBentryALTinterwordstretchfactor\fontdimen3\font minus
  \fontdimen4\font\relax}
\providecommand{\BIBforeignlanguage}[2]{{%
\expandafter\ifx\csname l@#1\endcsname\relax
\typeout{** WARNING: IEEEtran.bst: No hyphenation pattern has been}%
\typeout{** loaded for the language `#1'. Using the pattern for}%
\typeout{** the default language instead.}%
\else
\language=\csname l@#1\endcsname
\fi
#2}}
\providecommand{\BIBdecl}{\relax}
\BIBdecl

\bibitem{Maddah-Ali}
M.~A. Maddah-Ali and U.~Niesen, ``Fundamental limits of caching,'' \emph{IEEE
  Trans. Inf. Theory}, vol.~60, no.~5, pp. 2856--2867, 2014.

\bibitem{Golrezaei}
N.~Golrezaei, A.~F. Molisch, A.~G. Dimakis, and G.~Caire, ``Femtocaching and
  device-to-device collaboration: A new architecture for wireless video
  distribution,'' \emph{IEEE Commun. Mag.}, vol.~51, no.~4, pp. 142--149, Apr.
  2013.

\bibitem{Bastug}
E.~Bastug, M.~Bennis, and M.~Debbah, ``Living on the edge: The role of
  proactive caching in {5G} wireless networks,'' \emph{IEEE Communications
  Magazine}, vol.~52, no.~8, pp. 82--89, Aug. 2014.

\bibitem{Liu10}
H.~Liu, Z.~Chen, X.~Tian, X.~Wang, and M.~Tao, ``On content-centric wireless
  delivery networks,'' \emph{IEEE Wireless Communications}, vol.~21, no.~6, pp.
  118--125, Dec. 2014.

\bibitem{Tao2}
M.~Tao, E.~Chen, H.~Zhou, and W.~Yu, ``Content-centric sparse multicast
  beamforming for cache-enabled cloud ran,'' \emph{IEEE Trans. Wireless
  Commun.}, vol.~15, no.~9, pp. 6118--6131, Sept 2016.

\bibitem{Ao}
W.~C. Ao and K.~Psounis, ``Distributed caching and small cell cooperation for
  fast content delivery,'' in \emph{Proc. ACM MobiHoc}, Jun. 2015, pp.
  127--136.

\bibitem{Chen2}
Z.~Chen, J.~Lee, T.~Q.~S. Quek, and M.~Kountouris, ``Cooperative caching and
  transmission design in cluster-centric small cell networks,'' \emph{IEEE
  Trans. Wireless Commun.}, vol.~16, no.~5, pp. 3401--3415, May 2017.

\bibitem{Song}
J.~Song, H.~Song, and W.~Choi, ``Optimal caching placement of caching system
  with helpers,'' in \emph{Proc. IEEE ICC}, Jun. 2015, pp. 1825--1830.

\bibitem{Chae}
S.~H. Chae, J.~Y. Ryu, T.~Q.~S. Quek, and W.~Choi, ``Cooperative transmission
  via caching helpers,'' in \emph{Proc. IEEE Globecom}, Dec. 2015.

\bibitem{Xu3}
X.~Xu and M.~Tao, ``Modeling, analysis, and optimization of coded caching in
  small-cell networks,'' \emph{IEEE Trans. Commun.}, vol.~65, no.~8, pp.
  3415--3428, Aug. 2017.

\bibitem{Maddah-Ali3}
M.~A. Maddah-Ali and U.~Niesen, ``Cache-aided interference channels,'' in
  \emph{Proc. IEEE ISIT}, 2015, pp. 809--813.

\bibitem{Xu2}
F.~Xu, K.~Liu, and M.~Tao, ``Cooperative {Tx/Rx} caching in interference
  channels: A storage-latency tradeoff study,'' in \emph{Proc. IEEE ISIT}, Jul.
  2016, pp. 2034--2038.

\bibitem{Xu}
F.~Xu, M.~Tao, and K.~Liu, ``Fundamental tradeoff between storage and latency
  in cache-aided wireless interference networks,'' \emph{IEEE Trans. Inf.
  Theory}, vol.~63, no.~11, pp. 7464--7491, Nov. 2017.

\bibitem{Cao}
Y.~Cao, M.~Tao, F.~Xu, and K.~Liu, ``Fundamental storage-latency tradeoff in
  cache-aided {MIMO} interference networks,'' \emph{IEEE Trans. Wireless
  Commun.}, vol.~16, no.~8, pp. 5061--5076, Aug. 2017.

\bibitem{Cao2}
Y.~Cao, F.~Xu, K.~Liu, and M.~Tao, ``A storage-latency tradeoff study for
  cache-aided {MIMO} interference networks,'' in \emph{Proc. IEEE Globecom},
  Dec. 2016, pp. 1--6.

\bibitem{Bastug2}
E.~Bastug, M.~Bennis, M.~Kountouris, and M.~Debbah, ``Cache-enabled small cell
  networks: Modeling and tradeoffs,¡± , vol. 2015, no. 41, p. 2¨c11, feb.
  2015.'' \emph{EURASIP J. Wireless Commun. Netw.}, vol. 2015, no.~41, pp.
  2--11, Feb. 2015.

\bibitem{Shanmugam}
K.~Shanmugam, N.~Golrezaei, A.~G. Dimakis, A.~F. Molisch, and G.~Caire,
  ``Femtocaching: Wireless content delivery through distributed caching
  helpers,'' \emph{IEEE Trans. Inf. Theory}, vol.~59, no.~12, pp. 8402--8413,
  Dec. 2013.

\bibitem{Ali}
M.~M. Ali and M.~Obaidullah, ``Distribution of linear combination of
  exponential variates,'' \emph{Communications in Statistics - Theory and
  Methods}, vol.~11, no.~13, pp. 1453--1463, 1982.

\bibitem{David}
H.~A. David and H.~N. Nagaraja, \emph{Order Statistics}.\hskip 1em plus 0.5em
  minus 0.4em\relax John Wiley, 2003.

\bibitem{Hildebrand}
F.~B. Hildebrand, \emph{Introduction To Numerical Analysis}.\hskip 1em plus
  0.5em minus 0.4em\relax General Publishing Company, Ltd., 1987.

\bibitem{Ali1}
M.~M. Ali, ``Content of the frustum of a simplex,'' \emph{Pacific Journal of
  mathematics}, vol.~48, no.~2, pp. 313--322, 1973.

\end{thebibliography}

\end{document}